\documentclass{article}

\usepackage{arxiv}

\usepackage[utf8]{inputenc} 
\usepackage[T1]{fontenc}    
\usepackage{hyperref}       
\usepackage{url}            
\usepackage{booktabs}       
\usepackage{amsfonts}       
\usepackage{nicefrac}       
\usepackage{microtype}      
\usepackage{lipsum}
\usepackage{graphicx}
\usepackage{euler}
\usepackage{relsize}
\usepackage{braket}
\usepackage{amsmath,amssymb}
\numberwithin{equation}{section}
\graphicspath{ {./images/} }

\title{Ultrafast laser architectures for\\quantum control of nuclear fusion}

\author{
 Jake Levitt \\
  Cortex Fusion Systems, Inc.\\
  New York, NY 10128 \\
  \href{mailto:jakelevitt@cortexfusion.systems}{\texttt{jakelevitt@cortexfusion.systems}}} 

\begin{document}
\maketitle
\begin{abstract}
Quantum control of nuclear fusion involves engineering quantum coherences in a nuclear wavepacket to accelerate tunneling through the Coulomb barrier and modifying the analytic structure of the \textit{S}-matrix to facilitate long-range reactive capture. We present a three-body fusion reaction which is amenable to quantum control. The main result of the present inquiry is the discovery of an embodied class of ultrafast laser architectures [Levitt \textit{et al.}, U.S. Patent Application No. 17/855,476 (2022)] which realize solutions to the Schr\"odinger equation in a logically consistent manner as to that presented in [Saha \textit{et al.}, Mol. Phys. \textbf{110}, 9--10 (2012)]. Further, we provide some necessary (but not necessarily sufficient) conditions for net electrical power production using the optical designs here.
\end{abstract}


\section{Background}
\label{sec:quantum coherences}
\subsection{Quantum coherences: tunneling through \uppercase{C}oulombic barriers}
Progress in ultrafast laser science has delivered devices with high optical bandwidth which are capable of resonantly exciting a coherent superposition between many molecular states \cite{gregory2022laboratory}. In this scenario, molecular quantum dynamics are driven by both the populations of the molecular quantum states as well as their quantum coherences, leading to various non-classical phenomena in matter \cite{averbukh1989fractional,vrakking1996observation,kis1996entangled} which are, furthermore, controllable \cite{weinacht1999controlling,judson1992teaching}. In a model study of quantum dynamics, repeated $\pi$ phase-kicks to a basis state coupled with the continuum has been shown to modulate population transfer via control over the imaginary component of coherences responsible for quantum dynamics, at first order in perturbation theory. This yields unitary dynamics analogous to the projective quantum Zeno and anti-Zeno effects \cite{saha2011tunneling}. In the unitary case, the extent to which acceleration of population transfer (anti-Zeno effect) occurs is determined by the timing between successive operations of the phase-kick \cite{levitt2023coherent}. In 2012, a model of $\pi$ phase-kicks operating on a nuclear wavepacket (\textit{i.e.,} a coherent superposition of nuclear rotational and vibrational states) bound in a trap with Coulombic (MeV-scale) barriers was presented to achieve a highly accelerated tunneling escape into the continuum of fusion products with extremely high energy efficiency \cite{saha2012tunneling}. This offered the first evidence that dynamics modulated by the anti-Zeno effect could be useful in a coherent, ultrafast controlled fusion scenario involving an adequately prepared molecule. Specifically, the chemically bound nuclei must be coupled to narrow, near-threshold resonances of fusion products which have the same baryonic content as the molecule (\textit{e.g.,} a $^6Li(D,\alpha)^4He$ fusion reaction proceeding through the $2^+$ resonance in $^8Be$; the $2^+$ resonance energy is ($22.2+i0.8$)MeV and the $^6Li+D$ cluster breakup threshold in $^8Be$ is located at $22.2798$MeV above the ground state; $^6Li$ and $D$ are also found chemically bonded, so the states of the $^6LiD$ molecule are embedded as resonances in the $^8Be$ spectrum at the $^6Li+D$ cluster breakup threshold energy) \cite{belyaev1996fusion,belyaev1996possibility,belyaev1996perturbation}. A ``narrow, near-threshold resonance'' refers to a resonance in the composite nucleus which is located near a breakup threshold into charged clusters which can also chemically bind to form a molecule. The remainder of the present inquiry will involve discussions of the overlap between molecular and resonance states in this case, as well as the embodied mechanism which efficiently accesses the resonance starting from the molecular states. The key result of the present work is the presentation of a class of ultrafast laser designs which realize molecular dynamics in a logically consistent manner to that of the control protocol \cite{saha2012tunneling}, for a molecular system which also exhibits a non-zero, controllable overlap with a narrow, near threshold resonance in a composite nucleus formed from the baryonic components of the molecule. 
\subsection{Embodying phase-kick control with holonomy}
\label{sec:double slits}
Various embodiments of the $\pi$ phase-kick control have been proposed \cite{rego2007multiple,rego2009coherent} in which there is a one-to-one correspondence between optical pulses and phase-kicks. However in the model \cite{saha2012tunneling}, an ultrashort time between phase-kicks, of less than half a femtosecond (fs), was required to observe tunneling acceleration through a Coulombic barrier. This implies that repeated phase-kicks on a nuclear wavepacket must be executed by an individual fs pulse (\textit{i.e.,} there must be a one-to-many relationship between optical pulses and phase-kicks). In order to achieve this, we will consider molecular systems in which, additionally, an intrinsic feature of the electronic structure (the ability to induce degeneracies between two electronic surfaces with a laser) can be leveraged to realize repeated phase-kicks on a nuclear wavepacket according to the time spacing in \cite{saha2012tunneling} and accessible with one additional fs control pulse following ultrafast excitation. A nuclear wavepacket on $\mathbb{R}^2$ which completes a closed loop around a degeneracy between electronic surfaces acquires a geometric phase exactly equal to $\pi$ \cite{simon1983holonomy,berry1984quantal}. The acquisition of this geometric phase flips the sign of coherences between the basis states of the nuclear wavepacket and the narrow, near threshold resonance, leaving unaltered the projection of the total system wavefunction (molecule and composite nucleus) over the continuum of fusion product channels. The geometric phase associated with degeneracies between electronic surfaces is, therefore, the same $\pi$ phase-kick encountered in \cite{saha2011tunneling,levitt2023coherent,saha2012tunneling}. In general, degeneracies between two electronic surfaces equip the Hermitian line bundle over the nuclear Hilbert space (\textit{i.e.,} the projective Hilbert space of nuclear wavepacket configurations on a given electronic surface in the molecule), whose typical fiber is the $U(1)$ Lie group, with the holonomy of the structural group of the M\"obius strip fiber bundle $\mathbb{Z}_2\otimes U(1)$ \cite{ryabinkin2017geometric}. For $U(1)$ fibers, the holonomy manifests as a \textit{phase}: it can therefore be used for unitary (phase-kick) quantum control of the nuclear wavepacket. 

In essence, we will have the nuclear wavepacket execute its own phase-kicks simply by evolving on an electronic structure with degeneracies. The associated dynamics are furthermore amenable to ultrafast laser control as follows. In the Floquet picture of quantum mechanics, light-molecule interactions are described in terms of ``dressed states,'' molecular states that are shifted in energy according to the perturbational field frequency, intensity, and polarization. The dressed states determine effective light-induced potentials with the usual topological features of multidimensional potential energy surfaces, such as the emergence of light-induced electronic degeneracies \cite{videla2018floquet} which can inherit their positions from degeneracies in the natural electronic structure. In this regard, unitary control over the nuclear wavepacket with holonomy is effected by applying intense laser fields, to induce degeneracies between electronic surfaces at specific (desired) and time-dependent locations in the nuclear Hilbert space \cite{halasz2011conical,halasz2012light,halasz2012light2,halasz2013nuclear,halász2014influence,halasz2018geometric}. Control over the position, and emergent timing, of degeneracies in the light-renormalized electronic structure with a fs control pulse enables execution of repeated, ultrafast phase-kicks on the nuclear wavepacket as it evolves dynamically on the light-renormalized electronic structure following resonant excitation. In this regard, a two-pulse, phase-locked control protocol acting on the appropriate molecule will spawn a nuclear wavepacket and then develop a severe quantum interference pattern in the nuclear wavepacket, in correspondence with the control proposed by \cite{saha2012tunneling}. 

Separately, if the spectral width of the laser pulse used for control is smaller than that of the Frank-Condon manifold, but much larger than the vibrational frequency, then the nuclei move during the dynamics associated with ultrafast excitation, and the nuclear wavepacket is transformed into a squeezed state in which the quantum uncertainty of its momentum is larger than, \textit{i.a.,} that of the ground vibrational state of the corresponding vibrational potential \cite{janszky1986squeezing,janszky1990squeezing,janszky1994competition,dunn1993experimental,abrashkevich1994optimal,averbukh1993optimal,ohtsuki1998quantum,grigorenko2008analytical,cao1997simple,chang2005adiabatic,chang2006adiabatic}. The laser may therefore also be used to prepare a nuclear wavepacket with a large momentum spread around the energy of the narrow, near-threshold resonance, if the resonance is sufficiently low-lying with respect to the molecular states so as to be accessible with multiple photons provided by the control pulse. The trade-off here is between preparing the nuclear wavepacket in a configuration having a large overlap with the narrow, near threshold resonance and ionizing the molecule, which prohibits any further phase-kick control of the nuclear wavepacket.

We will see in \hyperref[sec:three-body system]{Subsection 1.5} and \hyperref[sec:simplest example device]{Section 2} that a molecule which fulfills conditions proposed in the previous paragraphs is the water molecule ($H_2O$), and the fusion reaction of interest is the $^{16}O(2p,\gamma)^{18}Ne$ reaction (\textit{i.e.,} two-proton capture by $^{16}O$).


\subsection{Nuclear \uppercase{v}ertex \uppercase{c}onstants: long-range reactive capture}
\label{sec:nvcs}
To get a sense of the coupling between molecular states and a narrow, near-threshold resonance -- starting from molecular states with energies in the deep sub-Coulomb regime -- we will employ the Nuclear Vertex Constants (NVCs) which are, for the $^{16}O(2p,\gamma)^{18}Ne$ fusion reaction, encoded explicitly by the overlap integral of \hyperref[eq:1]{Equation 1.5}. The NVCs specify the projection $\braket{\psi_{\textrm{mol}}^{JM}(X)|\psi_{\textrm{nuc}}^{JM}(X)}$ of the resonance wavefunction defined, \textit{e.g.,} by \hyperref[eq:4]{Equation 1.3}, onto the molecular wavefunction defined, \textit{e.g.,} by \hyperref[eq:3]{Equation 1.2}. We consider the nuclear wavepacket in a molecule to comprise a product of charged sub-clusters unbound with respect to the strong force, but which are chemically bound. The relationship between this projection and the NVCs proceeds via the Asymptotic Normalization Coefficients (ANCs) in Equation (3) and Equation (8) of \cite{mukhamedzhanov1999connection}. In the example system from the first paragraph, the sub-cluster wavefunction would specify the configuration of $^6Li$ and $D$ when they are chemically bound in the $^6LiD$ molecule, and the NVCs would provide the probability amplitude that the $2^+$ resonance in $^8Be$ is formed from the asymptotic tail of the  $^8Be$ nuclear wavefunction overlapping with the molecular wavefunction. The NVCs are necessary to consider in the current inquiry since the nuclear reaction of interest will proceed, because of the Coulomb barrier, at very large distances compared to the strong interaction radius between the clusters ($>>1$fm), and the NVCs determine entirely the cross section for such a long-range reactive capture (fusion) reaction \cite{burjan2020anc}. In the current inquiry we will furthermore see that in the special case of (\hyperref[sec:three-body system]{Subsection 1.5}), the NVCs are amenable to the squeezing control described in (\hyperref[sec:double slits]{Subsection 1.2}) via a renormalization of the breakup threshold energy in the field. The analysis (\hyperref[sec:three-body system]{Subsection 1.5}) provides grounding evidence, via computation of an overlap integral encoding the NVCs, that a long-range $^{16}O(2p,\gamma)^{18}Ne$ reactive capture reaction is possible.

\subsection{\textit{S}-matrix poles: engineering the width of narrow, near-threshold resonances}
\label{sec:s-matrix poles}
As we will see in \hyperref[sec:three-body system]{Subsection 1.5} for the $^{16}O(2p,\gamma)^{18}Ne$ fusion reaction, the width of the narrow, near-threshold resonance partially determines the NVCs, also as specified by Equation (32) of \cite{mukhamedzhanov1999connection}. In the present inquiry it becomes necessary to consider how to engineer this resonance width to facilitate the long-range reactive capture. We will proceed as follows: the \textit{S}-matrix $\mathbb{S}$ can be factored into Jost matrices \cite{ershov2021jost,rakityansky2013analytic,rakityansky2009generalized}
\begin{equation}
\mathbb{S}(k)=\frac{f^{(-)}(k)}{f^{(+)}(k)}    ,
\end{equation}
where $f^{(\pm)}(k)$ are the $\mathbb{K}$-space representation of the Jost functions and we have defined $\mathbb{S}$ in terms of its matrix elements. We consider the poles of $\mathbb{S}$ to be the zero points of $f^{(+)}(k)$, which correspond to bound states and resonances, and in turn depend on the quantum mechanical potential of the scattering process (\textit{e.g.,} the fusion reaction coordinate). It is then trivial to see how the analytic structure of $\mathbb{S}$ is amenable to control by strong fields. The narrow, near-threshold resonance is here interpreted as a complex pole of $\mathbb{S}$, in the second sheet of the analytic continuation of $\mathbb{S}$ to the complex plane of the momenta. In the Floquet picture of quantum mechanics, diverse $\mathbb{S}$ pole trajectories can be realized along the frequency \cite{landa2018singularities,li1999floquet}, intensity \cite{miyagi2010unified,miyagi2009unified,unnikrishnan1993semiclassical}, and polarization \cite{unnikrishnan1993semiclassical} degrees of freedom of an applied control pulse. The control pulse can also induce new poles \cite{magunov2001laser}, as well as change the location of branch points in the Riemann surface of the matrix elements of $\mathbb{S}$ \cite{martinez2001transmission}. The NVCs are directly related to the residue of $\mathbb{S}$ at the pole via the ANCs by Equation (8) of \cite{mukhamedzhanov1999connection} and Equation (40) of \cite{mukhamedzhanov2023resonances}. 

In general the NVCs are amenable to control by intense laser fields (\textit{See, e.g.,} \cite{martinez2001transmission} for controlling the residue of $\mathbb{S}$ at the pole). In the case of the $^{16}O(2p,\gamma)^{18}Ne$ reaction, transition probabilities starting from molecular states are sensitive to minor changes in the position of the pole for the narrow resonance near the breakup threshold into clusters (and in fact, there is an exponential sensitivity) (\hyperref[eq:1]{Equation 1.5}). Together these statements imply that the width of the narrow, near threshold resonance can be altered to modify the NVCs, and that, in the special scenario of (\hyperref[eq:1]{Equation 1.5}), the laser bandwidth and peak pulse intensity detailed in (\hyperref[sec:simplest example device]{Section 2}) are sufficient to effect control of the $^{16}O(2p,\gamma)^{18}Ne$ fusion reaction at the level of the analytic structure of $\mathbb{S}$. 
\subsection{Three-body system}
\label{sec:three-body system}
The water molecule ($H_2O$) is a platform on which the $^{16}O(2p,\gamma)^{18}Ne$ three-body fusion reaction \cite{belyaev1996fusion,belyaev1996possibility} can be assisted by making use of an ultrafast laser control protocol (\hyperref[sec:simplest example device]{Section 2}) which acts on water molecules. We recognize that in the traditional context of plasma at astrophysical temperatures and densities, three-body reactive collisions are exceedingly rare \cite{grigorenko2005three,marganiec2016experimental,casal2016radiative,parfenova2018coulomb,grigorenko2006soft,grigorenko2020asymptotic}. Following \cite{belyaev1996possibility}, we are going to show that in the case of the $^{16}O(2p,\gamma)^{18}Ne$ reaction starting from water, the cross section, as determined by an overlap integral encoding the NVCs for this reaction, is non-trivial and furthermore, amenable to control at the level of the analytic structure of $\mathbb{S}$ and specifically amenable to the control in \cite{saha2012tunneling}. Our control protocol acts on the nuclear wavepacket in a molecule and does not involve the creation of a plasma. In fact, it is a coherent process that occurs before thermalization, proceeding entirely according to Schr\"odinger dynamics. 

The $^{18}Ne$ spectrum exhibits a $3^+$ excited state resonance which is low-lying with respect to the three-body $p+p+^{16}$$O$ cluster breakup threshold \cite{bardayan1999observation,bardayan2000astrophysically,parpottas2005f} (\textit{i.e.,} the molecular states of water are embedded as resonances in the $^{18}Ne$ spectrum at the three-body $p+p+^{16}$$O$ cluster breakup threshold energy, which is close to the $3^+$ state), and $^{18}Ne$ is itself a halo nucleus even in the ground state, comprised of the bound clusters $p$, $p$, and $^{16}$$O$ \cite{lay2010three}. Following \cite{belyaev1996possibility}, we will now compute the overlap between the water molecule and the $3^+$ state in $^{18}Ne$ (\textit{i.e.,} encoding the NVCs for the $^{16}O(2p,\gamma)^{18}Ne$ reaction). We note that while in \cite{belyaev1996fusion,belyaev1996possibility} the $1^-$ state is identified as near-isoenergetic to the $p+p+^{16}O$ three-body breakup threshold energy in $^{18}Ne$, later more advanced spectroscopy of $^{18}Ne$ places the $3^+$ state nearly isoenergetic to the three-body breakup threshold \cite{bardayan1999observation,bardayan2000astrophysically,parpottas2005f} and the $1^-$ state relatively far below the threshold \cite{charity2019invariant}. For clarity we are taking the three-body breakup threshold energy reported in Figure (11) of \cite{charity2019invariant} ($4,523$keV above the $0^+$ ground state). In another commonly cited publication \cite{ajzenberg1987energy} the three-body breakup threshold is reported at $4,522$keV above the ground state; in either case the values of the $3^+$ state energy reported in \cite{bardayan1999observation,bardayan2000astrophysically} ($4,523.7\pm 2.9$keV) and \cite{parpottas2005f} ($4,527\pm 4$keV) are, within their respective systematic and statistical uncertainty of experiment, isoenergetic or nearly isoenergetic to the three-body breakup threshold.

Following \cite{belyaev1996possibility}, in order to compute the overlap between water and the $3^+$ resonance of $^{18}Ne$, we will employ an ansatz for the nuclear wavepacket (\textit{i.e.,} in the water molecule; \textit{cf.,} nuclear wavefunction) which accounts for the Coulomb repulsion between the $p+p+^{16}$$O$ clusters at small distances, as well as the separation between the $p+p+^{16}$$O$ clusters when they are chemically bound in the water molecule (\textit{i.e.,} accounting for the geometric size of the water molecule):
\begin{equation}
\label{eq:3}
\psi_{\textrm{mol}}^{JM}(X)=\frac{1}{N_{\textrm{mol}}}\frac{F_{5/2}(\eta_0,\kappa\rho)}{\rho^{5/2}} e^{-\kappa\rho}Y_{l\lambda}^{3M} (\hat{x},\hat{y}),  
\end{equation}
where the set of hyperspherical variables $X=\{\rho,\omega,\hat{x},\hat{y}\}$ has hyperradius $\rho=\sqrt{x^2+y^2}$ and hyperangle $\omega=\arctan{y/x}$. $\hat{x}=\{x,\theta_x\}$ and $\hat{y}=\{y,\theta_y\}$ are Jacobi variables. The Coulomb potential is $V(X)=\mathcal{V}(\Omega)/\rho$ with $\Omega=\{\omega,\hat{x},\hat{y} \}$ denoting the five angles in $X$. The regular solutions of the hyperradial Schr\"odinger equation are $F_{\nu}$ (the regular Coulomb wavefunctions), and $Y_{l\lambda}^{JM}(\hat{x},\hat{y})$ are eigenfunctions of the total angular momentum operator $\hat{J}=(\hat{l}+\hat{\lambda})+\hat{S}$, with $\hat{l}$ and $\hat{\lambda}$ being associated with $\hat{x}$ and $\hat{y}$, respectively, $\hat{S}$ the total spin of the $pp$ pair in the water molecule, and $M\in\left\{-J,-J+1,\dots,J\right\}$. $N_{\textrm{mol}}$ is a normalization factor and $\kappa=\sqrt{2\mu|\varepsilon_{\textrm{mol}}|}$ is the momentum corresponding to the chemical binding energy of the clusters in the water molecule ($\varepsilon_{\textrm{mol}}\approx-10$eV), with reduced mass $\mu=16m_p/33$. $\eta_0=\mathcal{V}_0/2\kappa$ is an effective Sommerfeld parameter, where $\mathcal{V}_0$ is obtained by averaging $\mathcal{V}(\Omega)$ with the angular part of $\psi_{\textrm{mol}}^{JM}(X)$. 

As an ansatz for the $3^+$ state (\textit{i.e.,} the wavefunction describing the composite $^{18}Ne$ nucleus; \textit{cf.,} nuclear wavepacket in the water molecule), we will employ the asymptotic form of the Coulombic three-body breakup function normalized to the nuclear volume:
\begin{equation}
\label{eq:4}
 \psi_{\textrm{nuc}}^{JM}(X)=\frac{1}{N_{\textrm{nuc}}}\frac{f^3(\rho,\omega)}{\rho^{5/2}}Y_{l\lambda}^{3M} (\hat{x},\hat{y}),  
\end{equation}
where
\begin{equation}
\label{eq:2}
f^J(\rho,\omega)=  \int \,d\hat{x}\,d\hat{y}  \exp\left\{iK\rho-i\frac{\mathcal{V}(\Omega)}{2K}\ln(2K\rho)\right\}Y_{l\lambda}^{JM}(\hat{x},\hat{y}),
\end{equation}
and $K\approx\sqrt{E}$ is the momentum corresponding to the outgoing $p+p+^{16}$$O$ clusters in a three-body breakup scenario starting from the $3^+$ state of $^{18}Ne$, or equivalently the gap between the total nuclear wavepacket energy (\textit{i.e.,} in the water molecule; \textit{cf.,} nuclear wavefunction) and the $3^+$ resonance energy. 

The computed overlap \cite{belyaev1996possibility} between these three clusters and the threshold $3^+$ resonance in $^{18}Ne$ (\textit{i.e.,} encoding the NVCs) is
\begin{equation}
\label{eq:1}
\mathscr{O}= \exp\left\{-\frac{\pi}{2} \eta_K^0\right\} \exp\left\{i \eta_K^0S\right\},    
\end{equation}
where $\eta_K^0=\mathcal{V}_{\textrm{min}}/2K$ is another effective Sommerfeld parameter with $\mathcal{V}_{\textrm{min}}$ the minimum value of the angular part $V(\Omega)$ of the total Coulomb potential. The phase $S$ depends on $\mathcal{V}_{\textrm{min}}$ and another parameter $\xi=K/\kappa$ which by construction has the range $0\leq\xi\leq\sqrt{\Gamma/|\varepsilon_{\textrm{mol}}|}$, with $\Gamma$ the partial width of the $3^+$ resonance in the two-proton channel. The two-proton partial width of a different, higher-lying $^{18}Ne$ resonance was measured experimentally in \cite{del2001decay,raciti2008experimental}, and is comparable to the magnitude of the chemical binding energy $|\varepsilon_{\textrm{mol}}|$. Considering that the one-proton partial width is the predominant contributor to the total width of the $3^+$ state \cite{bardayan1999observation,bardayan2000astrophysically,parpottas2005f}, and that the measured $\gamma$ partial width of the $3^+$ state is on the order of tens of meV \cite{bardayan2009direct,chipps2009first}, it is likely that the two-proton partial width of the $3^+$ state is of similar magnitude to that measured in \cite{del2001decay,raciti2008experimental} (\textit{i.e.,} it is extremely narrow), and therefore satisfies the conditions of \cite{belyaev1996perturbation}. There exists a wide subdomain of $S$ as a function of $\xi$ for which $\mathbb{I}m (S)<0$ and $|\mathbb{I}m (S)|>\frac{\pi}{2}$, which implies that the overlap integral (\hyperref[eq:1]{Equation 1.5}) can, or can be made to (\hyperref[sec:double slits]{Subsection 1.2}, \hyperref[sec:s-matrix poles]{Subsection 1.4}), increase exponentially with decreasing $K$. Starting from these three clusters bound by the electronic structure of water, and under the ultrafast laser control strategies described here, the $3^+$ state of $^{18}Ne$ can therefore be accessed from long range. 
\section{Optical design for facilitated tunneling and long-range reactive capture}
\label{sec:simplest example device}
The intramolecular quantum dynamics of water (\textit{i.e.,} the quantum dynamics of the nuclear wavepacket) that follow electronic excitation at the Lyman-$\alpha$ wavelength ($121$nm) are, furthermore, amenable to the phase-kick control described in (\hyperref[sec:double slits]{Subsection 1.2}) via the ``chemical double slits'' effect \cite{dixon1999chemical}. The effect is essentially a dynamical interference of the nuclear wavepacket with itself as it passes through two conical intersections (CIs; electronic degeneracies) belonging to the electronic structure of water (originating from degeneracies between the $\tilde{B}^1A_1$ and the $\tilde{X}^1A_1$ electronic states; on which the nuclear wavepacket is evolving in superposition) along the relaxation trajectory initiated by the exciting pulse, eliciting hundreds of nodes in the nuclear wavepacket. By renormalizing the electronic structure of water with a second control pulse, various additional light-induced CIs can be realized to further develop the nuclear wavepacket quantum interference pattern toward the reactivity described in \cite{saha2012tunneling}. 

\begin{figure}
  \centering
  \includegraphics[trim={1cm 0 1cm 0},width=\linewidth]{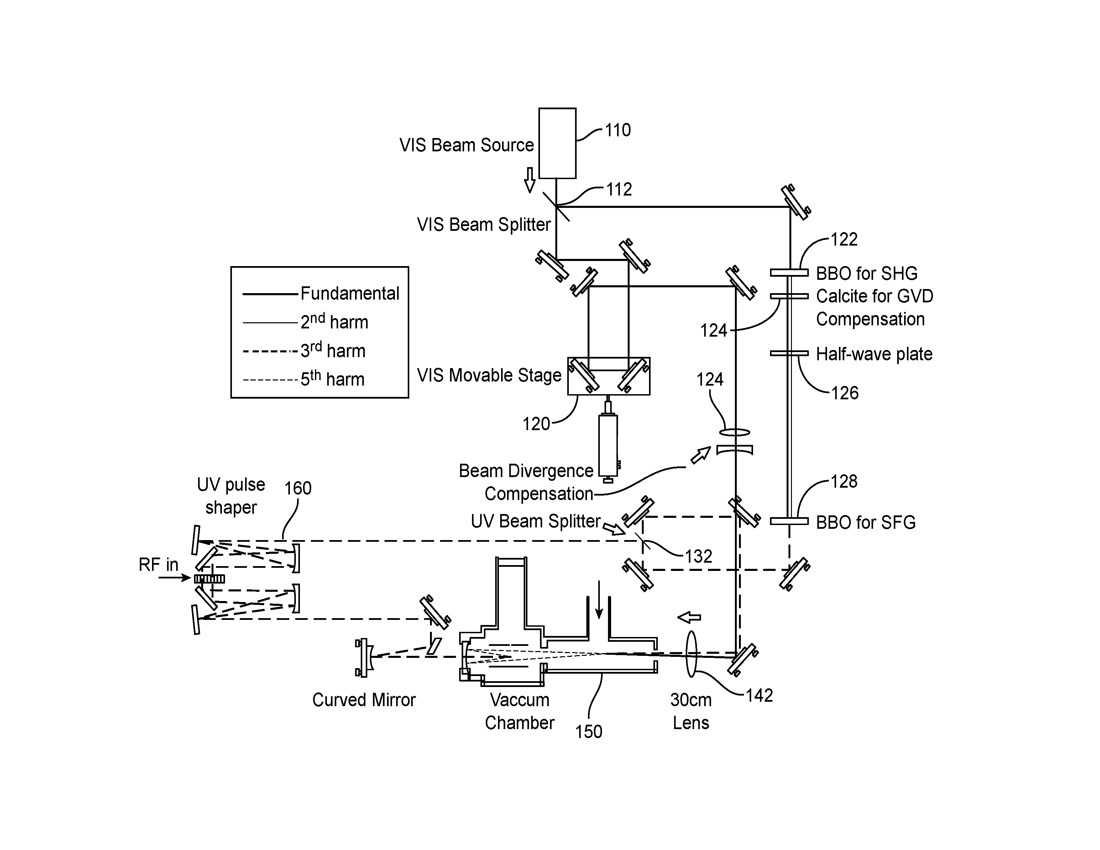}
  \caption{Configuration of the optical table for generation of low-intensity VUV light and moderate intensity DUV light, with phase-locking and fs-duration pulses.}
  \label{fig:fig1}
\end{figure}

We therefore conclude that the laser-assisted $^{16}O(2p,\gamma)^{18}Ne$ reaction starting from water requires a two-color, phase-locked ultrafast coherent control protocol, which prepares a nuclear wavepacket in water according to the ``chemical double slits'' effect, renormalizes the electronic structure of water to elicit reactive quantum interference patterns in the nuclear wavepacket according to \cite{saha2012tunneling}, and provides an optical bath of photons for pumping the nuclear wavepacket to, or closer to, the $3^+$ resonance energy (\textit{i.e.,} taking the $K\rightarrow0$ limit of (\hyperref[eq:2]{Equation 1.4}) \cite{belyaev1996possibility}) via multiphoton absorption by the nuclear wavepacket. A shaped, moderately intense field in the deep ultraviolet (DUV; on the order of $10^{12}$W/cm$^2$) interacts at a delay with the nuclear wavepacket in water prepared by a low-intensity vacuum ultraviolet pulse (VUV; on the order of $10^{11}$W/cm$^2$) to match, or nearly match, the total energy of the nuclear wavepacket with the $3^+$ resonance in $^{18}Ne$ (\textit{i.e.,} taking the $K\rightarrow0$ limit of (\hyperref[eq:2]{Equation 1.4}) \cite{belyaev1996possibility}). The pulses also induce multiphoton absorption from, or emission to, the field, by the molecule \cite{chu2004beyond}, to create CIs in the laser-renormalized electronic structure (consider, \textit{e.g.,} light-induced CIs originating from degeneracies among the light-induced potentials $\tilde{D}^1A_1-\hbar\omega$ and $\tilde{X}^1A_1+\hbar\omega$, and the $\tilde{B}^1A_1$ electronic state; all occurring below the first vertical ionization energy of water), which act as a highly augmented ``chemical double slits'' \cite{dixon1999chemical} and endow the nuclear wavepacket with a severe interference pattern directly as a consequence of repeated phase-kicks as discussed in (\hyperref[sec:double slits]{Subsection 1.2}). Combined these two pulses yield a sharp deviation from the exponential decay into fusion products expressed in \cite{belyaev1996perturbation}, concordantly with the anti-Zeno mechanism of \cite{saha2012tunneling}. The delicate interplay of both multiphoton effects, guided by the delay stage \{120\} and the DUV pulse shaper \{160\} configurations of Figure \ref{fig:fig1}, is required to observe a reactive quantum interference pattern having the appropriate average energy and energy spread to match a narrow, near-threshold resonance. This necessitates bichromatic, phase-locked pulses of mode-locked radiation, which are responsible for pumping a prepared nuclear wavepacket in fs-scale simultaneity to developing interferences.

$^{18}Ne$ decay products (\textit{e.g.,} via the cascade $\gamma$ decay $M2$($9.7$keV, $3^+$ $\rightarrow$ $1^-$) and $E1$($4,514$keV, $1^-$ $\rightarrow$ $0^+$) \cite{bardayan1999observation,bardayan2000astrophysically,charity2019invariant}, followed by additional $\beta^+$ decays from the ground state) can then be captured and used to generate electricity. 

\begin{figure}
  \centering
  \includegraphics[trim={1cm 0 1cm 0},width=\linewidth]{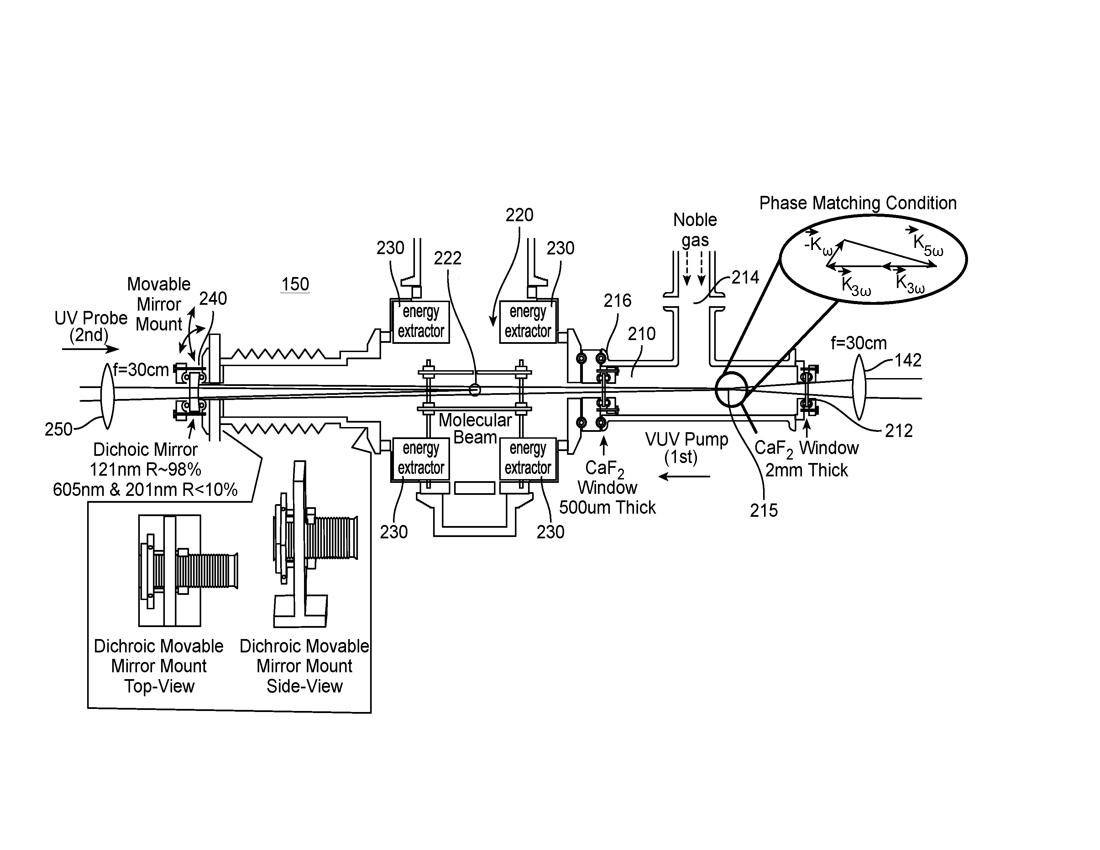}
  \caption{Configuration of the reaction chamber for bichromatic optical control of quantum tunneling with fuel deposited to the focal volume via molecular beam, and creation of light at the Lyman-$\alpha$ wavelength via non-collinear four-wave mixing.}
  \label{fig:fig2}
\end{figure}

In Figure \ref{fig:fig1} and Figure \ref{fig:fig2} we present an ultrafast laser architecture based on the bichromatic control protocol \cite{mypatent}. The key elements of the optical design are: initialization of the nuclear wavepacket in water under conditions observed in \cite{dixon1999chemical} (the ``chemical double slits'') with a first pulse at the Lyman-$\alpha$ wavelength ($121$nm), and subsequently, renormalization of the electronic structure of water with a second pulse of a different color to generate a severe quantum interference pattern in the nuclear wavepacket which exhibits thousands of nodes in compelling agreement with the control described by \cite{saha2012tunneling} and realized on the fs timescale with holonomy. The two pulses also prepare the nuclear wavepacket in water at, or near, the $3^+$ resonance energy of $^{18}Ne$ (\textit{i.e.,} taking the $K\rightarrow0$ limit of (\hyperref[eq:2]{Equation 1.4}) \cite{belyaev1996possibility}). Achieving control over interferences in a parsimonious manner (\textit{i.e.,} via the photonics described here) lends favorable energy balance to quantum-controlled fusion.

\subsection{Configuration of the optical table}
\textbf{Figure \ref{fig:fig1}}: In the embodiment shown, the fusion system includes a laser source \{110\} and a reactor \{150\}. Various optical components are used to split and modify the beam generated by the laser source \{110\} and direct the resulting beams into the reactor \{150\}. In other embodiments, the fusion system may include different or additional elements. Furthermore, various elements may operate in a different manner than described. The described fusion system is provided by way of examples of the broader principles it embodies. The laser source \{110\} generates an optical beam having a fundamental frequency. The laser source \{110\} includes a Kerr-lens mode-locked oscillator, pumped with a Coherent V5 continuous wave laser, and a multi-pass ring cavity amplifier, pumped with a Photonics DM-20 Q-switched $170$ns $Nd$:$YLF$ laser. The output of the amplifier may be a $1.5$mJ pulse with $1$kHz repetition rate, $30$fs pulse duration, and a central wavelength of $780$nm. The pulses may be spectrally broadened in a hollow-core fiber and the blue side of the spectrum (\textit{e.g.,} $605$nm) may be selected. A beam-splitter \{112\} splits the beam into two portions. The first portion is used for generating a DUV beam and the second portion is used for generating a VUV beam. The first portion of the fundamental frequency beam is directed to a first $\beta$-barium borate ($BBO$) crystal \{122\}. The $BBO$ crystal \{122\} is $250\mu$m-thick and cut for second harmonic generation (\textit{e.g.,} $\theta=40.3^{\circ}$, $\phi=0^{\circ}$, Type I SHG). The resulting intermediate second harmonic beam may pass through a calcite plate \{124\} (\textit{e.g.,} a $1$mm-thick calcite crystal, $\theta=41^{\circ}$, $\phi=0^{\circ}$) to compensate for group velocity dispersion (GVD) and a half-wave plate \{126\} to make the fundamental and second harmonic the same polarization. The fundamental and second harmonic beams may be used to generate one or more additional beams of different harmonics. The fundamental and second harmonic beams are combined in a second $100\mu$m-thick $BBO$ crystal \{128\} cut for third harmonic generation ($\theta=76^{\circ}$, $\phi=0^{\circ}$) in order to generate light at $201$nm via sum-frequency generation (SFG). After SFG $BBO$ crystal \{128\}, high-reflectivity dielectric mirrors for the third harmonic may be used to separate out the third harmonic from the fundamental and second harmonic. A second beamsplitter \{132\} may be used to split the third harmonic beam into two portions. The second beamsplitter \{132\} may be an uncoated $1$mm-thick $CaF_2$ window at $45^{\circ}$ to the beam. The first portion of the third harmonic beam is directed to a first optical input of the reactor \{150\} while the second portion is directed to a second optical input of the reactor via a DUV pulse shaper \{160\}. The first portion of the third harmonic beam is directed to a lens \{142\} (\textit{e.g.,} a $30$cm $CaF_2$ lens) that focuses the DUV light of the third harmonic beam and the second portion of the the beam having the fundamental frequency into a noble gas cell (\textit{e.g.,} within the reactor \{150\}) to generate a fifth harmonic VUV beam. A telescope or other optical system may be used to correct for the chromatic aberrations induced by the lenses \{142, 250\}. In one embodiment, generation of the fifth harmonic (\textit{e.g.,} $121$nm) is achieved by non-collinear four-wave mixing in noble gas (\textit{e.g.,} argon) that satisfies the following phase-matching condition:
\begin{equation}
 \vec{k}_{5\omega}=2\vec{k}_{3\omega}-\vec{k}_{\omega}  
\end{equation}
The pressure of the noble gas cell and the phase-matching angles of the pulses may be calibrated experimentally. 


The DUV pulse shaper \{160\} \cite{pearson2007shaped} manipulates the second portion of the third harmonic beam into control pulses that generate light-induced CIs in the fuel which develop quantum interference patterns realizing the unitary phase-kick control of tunneling \cite{saha2012tunneling}. A radiofrequency (RF) signal is sent to an acousto-optic modulator to generate a sound wave from which the optical pulses are diffracted. This enables modifying the phase and amplitude of the different optical frequencies to shape the DUV pulse. The RF signal can be modulated to shape the optical pulses and obtain a desired pulse shape. Pattern-recognition or machine learning models may be used to identify the appropriate pulse shapes which optimally realize phase-kick control with a high fusion output. In embodiments, ``closed loop learning control'' is used. A measurement of the yield is performed for a collection of random pulse shapes/sequences. A collection of the best pulse shapes is used to start a search for an optimal pulse shape/sequence using a pattern recognition algorithm. Given that one expects a very low fusion yield for a random initial pulse, the system may be initialized with some pulse shapes that are good first guesses given prior knowledge/experience. Any suitable observable indicative of the amount of fusion occurring may be used to help guide the system close to the ultimate goal. 
\subsection{Reaction chamber}
\textbf{Figure \ref{fig:fig2}}: In the embodiment shown, the reactor \{150\} includes a noble gas cell \{210\} and a reaction chamber \{220\}. The noble gas cell \{210\} holds a noble gas or a mixture of noble gasses at a predetermined pressure to facilitate generation of the fifth harmonic beam. The reaction chamber is where the pulsed optical beams interact with the fuel to facilitate fusion. In other embodiments, the reactor \{150\} may be configured differently. The noble gas cell \{210\} is supplied noble gas via a gas inlet \{214\} and maintained at a predetermined pressure. In embodiments, a few hundred Torr, a pressure between approximately $0.1$ and $1$ atmosphere ($76$ to $760$ Torr) is used. In one embodiment, the third harmonic and fundamental frequency beams enter the noble gas cell \{210\} through an optical input \{212\} (\textit{e.g.,} a $2$mm-thick $CaF_2$ window). The beams interact via non-collinear four-wave mixing to generate a fifth harmonic beam. Insert \{215\} illustrates an example phase-matching condition for generation of the fifth harmonic beam. The fifth harmonic beam passes through an optical output \{216\} (\textit{e.g.,} a $500\mu$m-thick $CaF_2$ window), which is a second optical input into the reaction chamber \{220\}. The reaction chamber \{220\} includes a molecular nozzle \{222\} configured to generate a molecular beam of the fuel. In Figure \ref{fig:fig2}, the molecular beam is directed to be coming directly out of the page. The reaction chamber may be maintained at low pressure (\textit{e.g.,} $10$$^{-7}$ Torr). In one embodiment, the VUV pulse is then reflected by a dichroic mirror \{240\} (\textit{e.g.,} with a radius of curvature $R=268$mm). The mirror may have a high reflectivity coating of $>90$\% at $0^{\circ}$ for $121$nm light and $<5$\% reflectivity for $201$nm and $605$nm. This enables the residual DUV and visible radiation left over from VUV generation to be separated from the VUV. The reflected VUV pulse is focused over the molecular nozzle \{222\}. The DUV reserved for the second pulse is sent through the dichroic mirror \{240\} and also focused over the molecular nozzle \{222\} with a lens \{250\} (\textit{e.g.,} a $30$cm $CaF_2$ lens). The VUV pulse reflected from the dichroic mirror \{240\} inside the reaction chamber \{220\} is steered over the molecular nozzle \{222\}. In one embodiment, this is done using a movable mirror mount. An example movable mirror mount is shown in the inset \{270\} of the figure. The illustrated movable mirror mount includes a KF40 blank with a hole drilled through the center of an O-ring groove set around the hole as a window holder. The KF40 window holder is connected to a KF40 bellow. Around the neck of the bellow collar there are three holes in the side where the ball tip head of a high precision 1/4''-80 Fine Hex Adjuster 20 can sit. An aluminium (or other suitable material) frame with three slots is positioned to lock 1/4''-80 Locking Bushings with Nuts in place. This aluminium frame may be bolted in place independent of the reaction chamber \{220\}. The 1/4''-80 Fine Hex Adjusters are threaded through the 1/4''-80 Locking Bushings. When the reaction chamber \{220\} comes under vacuum, the bellow contracts and the 1/4''-80 Fine Hex Adjuster ball tip heads come into contact with the holes in the collar. This acts like a Gimbal mount for the dichroic mirror \{240\} and enables steering of the VUV beam. The fifth harmonic and third harmonic pulses interact with fuel molecules in the molecular beam to facilitate fusion via the control protocol described here. The reaction chamber also includes one or more energy extractors \{230\}. The energy extractors \{230\} interact with fusion products to extract energy. For example, a high-efficiency scintillator/semiconductor system may be used to convert the fusion products into visible light and then electrical energy, \textit{etc.} It should be appreciated that any suitable energy extraction methods may be used to capture the energy released by fusion reactions. Fusion products released isotropically from the focal volume in the interaction region of the reactor \{150\} are collected by energy extractors typically configured to maximize the surface area of the energy extractors around the full $4\pi$ solid angle within the confines imposed by other elements of the system, at a distance according to the damage threshold of the scintillators, the radiant intensity of fusion products, and radiation transport properties of the scintillator.
\section{Conclusion: necessary (but not necessarily sufficient) conditions for net electrical power production}
Assuming a $<20$\% scintillator efficiency and a semiconductor system at the Shockley-Quiesser limit of efficiency (about $30$\%), the energy available as electrical current from the $^{16}O(2p,\gamma)^{18}Ne$ fusion reaction is on the order of a few hundred keV (a few million kJ/mol of water), for the decay chain presented at the beginning of \hyperref[sec:simplest example device]{Section 2}. If one starts with laser pulses having an energy of $1$mJ, assuming a $1$-$3$\% laser wall-power efficiency, this requires on the order of $10^{12}$ fusion events per laser pulse in order to reach parity between input and output electrical power. Supersonic molecular beams can yield number densities on the order of $10^{16}$ molecules per cc (and above) \cite{even2015even}. With a laser focus of $100$$\mu$m, and a VUV and DUV pulse propagation path length of $1$cm in the gas phase, the total focal volume is $10^{-4}$cc, so the total number of molecules in the focal volume can be about $10^{12}$ molecules (and above). With a VUV conversion efficiency of $0.01$\% \cite{horton2017ultrafast}, one can produce on the order of $10^{12}$ VUV photons per laser pulse, and if all these photons are absorbed then a fusion yield of near-unity under control achieves net electrical power production. By engineering higher VUV conversion and laser wall-power efficiencies, lower fusion yields become required to reach net electrical power production. Considering the tunneling efficiencies accessible under coherent control \cite{saha2012tunneling} and the overlap \cite{belyaev1996possibility}, there is reasonable optimism that such a net electrical power production is possible with contemporary bandwidth technologies. 





\end{document}